\documentclass{article}
\usepackage{spconf,amsmath,graphicx}

\newlength\savedwidth
\newcommand{\wcline}[1]{\noalign{\global\savedwidth\arrayrulewidth\global\arrayrulewidth 1.0pt} \cline{#1}
\noalign{\global\arrayrulewidth\savedwidth}}
\usepackage{multirow}
\usepackage{amsmath}
\usepackage{amsfonts}
\usepackage{colortbl}
\usepackage{cite}
\usepackage{arydshln}
\usepackage{url}

\newcommand{\bm}[1]{{\mbox{\boldmath $#1$}}}

\makeatletter
\def\bstctlcite{\@ifnextchar[{\@bstctlcite}{\@bstctlcite[@auxout]}}
\def\@bstctlcite[#1]#2{\@bsphack
\@for\@citeb:=#2\do{%
\edef\@citeb{\expandafter\@firstofone\@citeb}%
\if@filesw\immediate\write\csname #1\endcsname{\string\citation{\@citeb}}\fi}%
\@esphack}
\makeatother

\usepackage{color}

\title{Sound event detection guided by semantic contexts of scenes}
\name{\begin{tabular}{c}Noriyuki Tonami$^{1}$, Keisuke Imoto$^{2}$, Ryotaro Nagase$^{1}$, \\
Yuki Okamoto$^{1}$, Takahiro Fukumori$^{1}$, Yoichi Yamashita$^{1}$\end{tabular}}
\address{$^1$Ritsumeikan University, Japan, $^2$Doshisha University, Japan}

\begin{document}
\ninept
\maketitle
\begin{abstract}
Some studies have revealed that contexts of scenes (e.g., ``home,'' ``office,'' and ``cooking'') are advantageous for sound event detection (SED).
Mobile devices and sensing technologies give useful information on scenes for SED without the use of acoustic signals.
However, conventional methods can employ pre-defined contexts in inference stages but not undefined contexts.
This is because one-hot representations of pre-defined scenes are exploited as prior contexts for such conventional methods.
To alleviate this problem, we propose scene-informed SED where pre-defined scene-agnostic contexts are available for more accurate SED.
In the proposed method, pre-trained large-scale language models are utilized, which enables SED models to employ unseen semantic contexts of scenes in inference stages.
Moreover, we investigated the extent to which the semantic representation of scene contexts is useful for SED.
Experimental results performed with TUT Sound Events 2016/2017 and TUT Acoustic Scenes 2016/2017 datasets show that the proposed method improves micro and macro F-scores by 4.34 and 3.13 percentage points compared with conventional Conformer- and CNN--BiGRU-based SED, respectively.

\end{abstract}
\vspace{-3pt}
\vspace{-10pt}
\section{Introduction}
\label{sec:intro}
\vspace{-5pt}
The analysis of various environmental sounds in real life has been attracting significant attention \cite{Imoto_AST2018_01}.
The automatic analysis of various sounds enables many applications, such as anomaly detection systems \cite{koizumi_taslp2019}, life-logging systems \cite{Stork_ROMAN2012_01}, and monitoring systems \cite{Ntalampiras_ICASSP2009_01}.

Sound event detection (SED) is the task of estimating sound events (e.g., bird singing, footsteps, and wind blowing) and their time boundaries from acoustic signals.
In SED, a number of neural-network-based approaches have been proposed, including the convolutional neural network (CNN) \cite{Hershey_ICASSP2017_01}, recurrent neural network (RNN) \cite{Hayashi_TASLP2017_01}, and convolutional recurrent neural network (CRNN) \cite{SED_CRNN}.
CNN automatically extracts features and RNN models temporal structures.
CRNN is a hybrid of CNN and RNN, which is widely used as a baseline system of SED. 
More recently, to handle longer sequences, non-autoregressive models, such as Transformer \cite{transformer_kong,transformer_miyazaki} and Conformer \cite{conformer_miyazaki}, have been studied. 

Furthermore, some studies have revealed that the contexts of scenes (e.g., ``home,'' ``office,'' and ``cooking''), which are defined by locations, activities, and time, help increase the accuracy of SED \cite{Mesaros_EUSIPCO2011_01,Heittola_JASM2013_01,helen_interspeech2019,imoto_tonami_icassp2020,Komatsu_icassp2020,tonami_IEICE2021,dcasenet_icassp2021,tonami_icassp2021}.
For example, Heittola $\textit{et al}$. \cite{Heittola_JASM2013_01} have proposed a cascade method for SED using results of acoustic scene classification (ASC). 
Bear $\textit{et al}$. \cite{helen_interspeech2019}, Tonami $\textit{et al}$.\textcolor{black}{\cite{tonami_IEICE2021},} and Komatsu $\textit{et al}$. \cite{Komatsu_icassp2020} have proposed joint models of SED and ASC to take advantage of the relationships between sound events and scenes; e.g., the sound event ``mouse wheeling'' tends to occur in the scene ``office,'' whereas the event ``car'' is likely to occur in the scene ``city center.''
Cartwright $\textit{et al}$. \cite{Cartwright2020} have proposed audio tagging considering spatiotemporal contexts such as a city block, an hour, and a day.

\begin{figure}[t!]
  \centering
  \includegraphics[width=0.68\columnwidth]{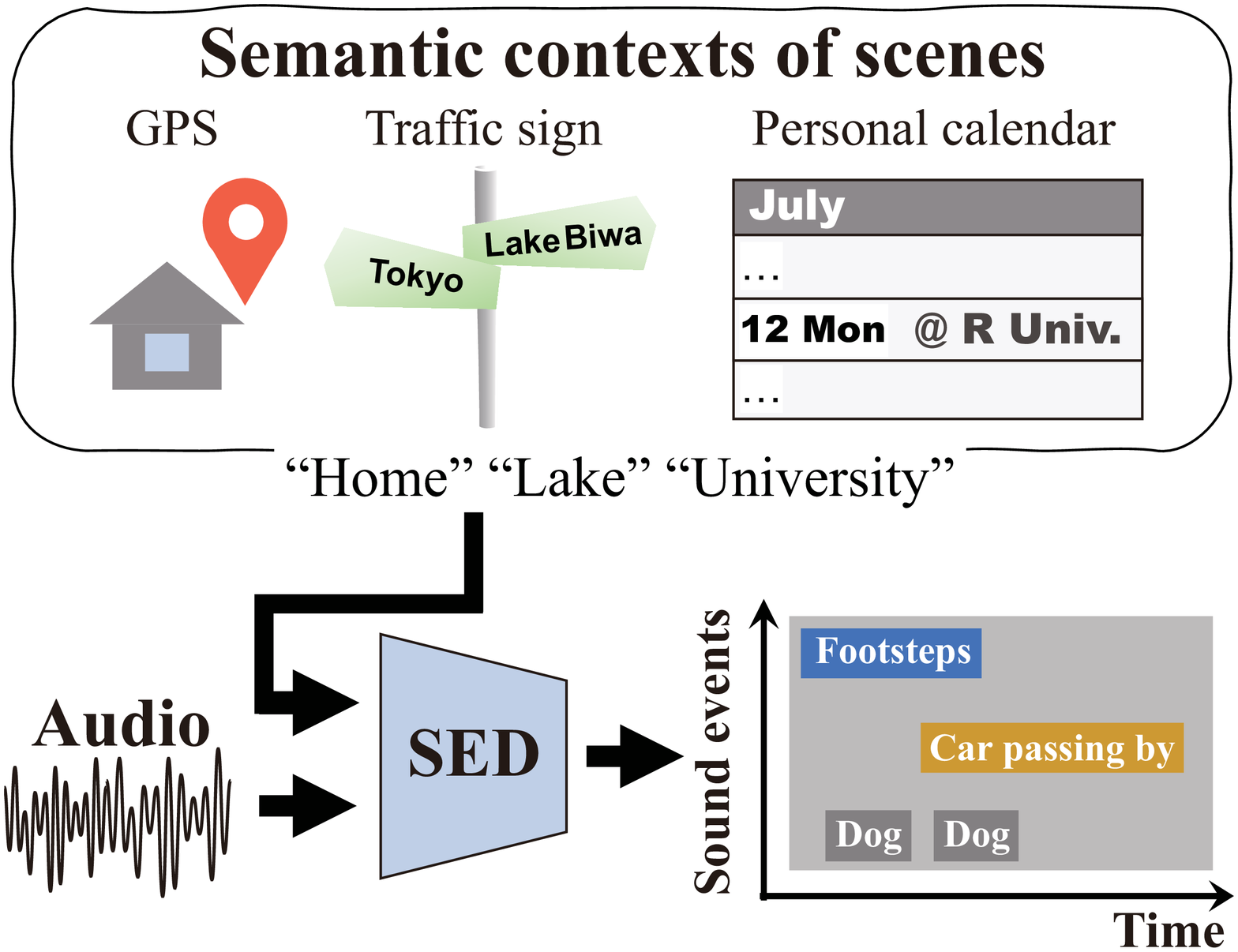}
  \vspace{-14pt}
  \caption{Overview of proposed method with semantic contexts of scenes}
  \label{fig:example_semantic_info_of_scene}
  \vspace{-10pt}
\end{figure}

Mobile devices (e.g., smartphones) and sensing technologies (e.g., global positioning system (GPS)) provide useful information on scenes for SED without the use of acoustic signals, as shown in Fig. \ref{fig:example_semantic_info_of_scene}.
However, conventional SED methods  \cite{Heittola_JASM2013_01,Komatsu_icassp2020,Cartwright2020} can utilize pre-defined contexts in inference stages but not  undefined contexts.
This is because one-hot representations or identifiers of pre-defined scenes are used as prior contexts for such conventional methods.

To mitigate this problem, we propose scene-informed SED where pre-defined scene-agnostic contexts are available for precisely detecting sound events.
To this end, pre-trained large-scale language models are utilized, which enables SED models to use unseen semantic contexts of scenes in inference stages.
To extract better scene contexts for SED, we further verify the effectiveness of the semantic representation of scene contexts, i.e., representation learning.

\begin{figure}[t!]
  \centering
  \includegraphics[width=0.68\columnwidth]{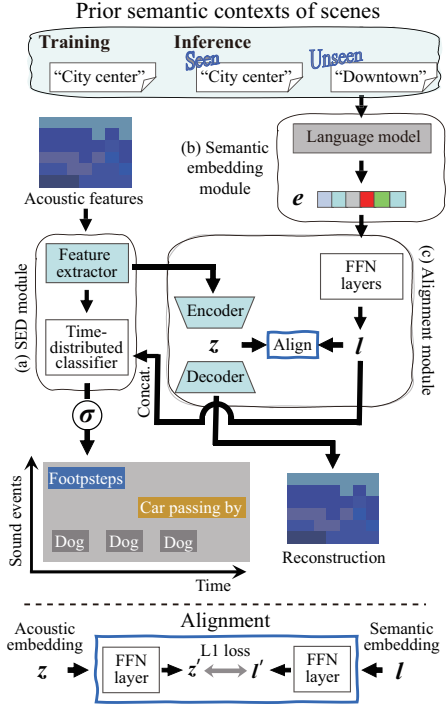}
  \vspace{-10pt} 
  \caption{Network architecture of proposed method}
  \label{fig:example_propesed_network}
  \vspace{-5pt}
\end{figure}

%---------------------------------
\vspace{-5pt}
\section{Framework of SED}
\label{sec:conv}
\vspace{-5pt}
\textcolor{black}{In SED, the goal is to predict active (1) or inactive (0) sound events; $\hat z_{n,t} \in \{0, 1\}^{N \times T}$ for each sound event $n$ and time frame $t$ from given acoustic features $x_{t,f}$.
Here, $N$ and $T$ are the total numbers of sound event classes and time frames, respectively.}
In general, $x_{t,f}$ is an element of acoustic features represented by the time-frequency domain such as log-mel spectrograms, where $f$ is the index of a dimension of the acoustic features.
In recent SED, neural-network-based methods have been dominant.
\textcolor{black}{In this work, we focus on strongly supervised SED, where ground truths of time stamps are available for the training.}
The binary cross-entropy is used to optimize neural-network-based SED models as follows: 

\vspace{-15pt}   
\begin{align}
\hspace*{0pt} {\mathcal L}_{\rm SED} 
 = - \! ~\frac{1}{NT}\sum^{N}_{n=1} \sum^{T}_{t=1} \! {\Big \{} z_{n,t} \log {\sigma}(y_{n,t}) \! 
\nonumber\\[-5pt]
+ ~\! (1 \! - \! z_{n,t}) \log & {\big (} 1 \! - \! {\sigma}(y_{n,t}) {\big )} {\Big \}},
\label{eq:event_loss}
\vspace{-35pt}
\end{align}
\vspace{-13pt}

\noindent where $z_{n,t} \in \{ 0,1 \}$ is a target label of a sound event, $y_{n,t} \in [0, 1] $ indicates the output of the SED network, and $\sigma(\cdot)$ denotes the sigmoid function.
In inference stages, $y_{n,t}$ is binarized using a pre-defined threshold value to obtain $\hat z_{n,t}$.

\setcounter{table}{0}
\begin{table*}[]
	\small
	\centering
	\caption{Overall results for SED. $\textless$$\tt 1$$\textgreater$ to $\textless$$\tt 6$$\textgreater$ are conventional methods. $\textless$$\tt 7$$\textgreater$ to $\textless$$\tt 14$$\textgreater$ are proposed methods.}
	\vspace{0pt}
	\label{tbl:overall_results}
	\scalebox{0.92}[0.92]{
		\begin{tabular}{llccll}
			\wcline{1-6}
			& & & & \\[-8pt]
			\multicolumn{1}{l}{Method} & \multicolumn{1}{l}{SED module} & \multicolumn{1}{c}{\it How scene is represented } & \multicolumn{1}{c}{\it How scene is fed to SED module} & \multicolumn{1}{c}{Micro F-score} & \multicolumn{1}{c}{Macro F-score}\\\hline 
			& & & & \\[-9pt]
			\tt{$\textless$1$\textgreater$} & CNN--BiGRU \cite{SED_CRNN} & - & - & 44.04\% $\pm{2.11}$ & 11.64\% $\pm{1.23}$ \\
			\tt{$\textless$2$\textgreater$} & MTL of SED \& ASC \cite{tonami_IEICE2021} & - & - & 44.76\% $\pm{1.63}$ & 11.70\% $\pm{1.13}$\\ 
			\tt{$\textless$3$\textgreater$} & MTL of SED \& SAD [31] & - & - & 45.22\% $\pm{1.58}$ & 11.81\% $\pm{1.23}$\\ 
			\tt{$\textless$4$\textgreater$} & Conformer \cite{conformer_miyazaki} & - & - & 47.23\% $\pm{2.79}$ & 11.95\% $\pm{0.61}$ \\
			\tt{$\textless$5$\textgreater$} & CNN--BiGRU & One-hot & Direct concatenation & 47.04\% $\pm{2.45}$ & 13.82\% $\pm{1.25}$ \\
			\tt{$\textless$6$\textgreater$} & MTL of SED \& ASC \cite{Komatsu_icassp2020} & One-hot & Direct concatenation & 46.92\% $\pm{3.62}$ & 13.40\% $\pm{0.83}$ \\\hdashline
			& & & & \\[-9pt]
			\tt{$\textless$7$\textgreater$} & CNN--BiGRU & BERT embedding & Direct concatenation & 46.91\% $\pm{2.13}$ & 13.79\% $\pm{1.22}$ \\
			\tt{$\textless$8$\textgreater$} & CNN--BiGRU & GPT2 embedding & Direct concatenation & 46.79\% $\pm{2.21}$ & 13.34\% $\pm{1.06}$ \\
			\tt{$\textless$9$\textgreater$} & CNN--BiGRU & One-hot & Aligned concatenation & 46.58\% $\pm{1.76}$ & 14.18\% $\pm{0.94}$ \\
			\tt{$\textless$10$\textgreater$} & MTL of SED \& ASC & One-hot & Aligned concatenation & 48.28\% $\pm{0.99}$ & 13.79\% $\pm{0.84}$ \\
			\tt{$\textless$11$\textgreater$} & CNN--BiGRU & BERT embedding & Aligned concatenation & 48.00\% $\pm{1.84}$ & 14.16\% $\pm{1.02}$ \\
			\tt{$\textless$12$\textgreater$} & CNN--BiGRU & GPT2 embedding & Aligned concatenation & 48.17\% $\pm{1.98}$ & \bf 14.77\% $\pm{0.82}$ \\
			\tt{$\textless$13$\textgreater$} & Conformer & BERT embedding & Aligned concatenation & \bf 51.57\% $\pm{2.62}$ & 12.76\% $\pm{0.74}$ \\
			\tt{$\textless$14$\textgreater$} & Conformer & GPT2 embedding & Aligned concatenation & 50.03\% $\pm{2.25}$ & 12.90\% $\pm{0.68}$ \\\wcline{1-6}
			& & & & \\[-9pt]
		\end{tabular}
	}
	\vspace{-15pt}
\end{table*}

\setcounter{table}{1}
\vspace{-15pt}
\begin{table}[t]
	\small
	\caption{Experimental conditions}
	\vspace{0pt}
	\label{tbl:parameter}
	\centering
	%\hspace{40pt}
	\scalebox{0.89}[0.89]{
		\begin{tabular}{ll}
			\wcline{1-2}
			&\\[-7pt]
			\textbf{SED module}\\
			& \\[-9pt]
			Network architecture & 3 CNN + 1 BiGRU + 1 FFN\\
			\# channels of CNN layers & 128, 128, 128 \\
			Filter size \textcolor{black}{($T\times F$)}& 3$\times$3 \\
			Pooling size \textcolor{black}{($T\times F$)}& 8$\times$1, 2$\times$1, 2$\times$1 (max pooling) \\
			&\\[-9pt]
			\# units in BiGRU layer & 32 \\
			\# units in FFN layers & 128, 48\\
			\# units in output layer & 25 \\\hdashline
			&\\[-9pt]
			\textbf{CNN--AE in alignment module}\\
			& \\[-9pt]
			\multirow{2}{*}{Network architecture} & 1 CNN (encoder) + \\
			& 3 Deconvolution (decoder)\\
			\# channels of CNN layers & 64, 128, 128, 128 \\
			Filter size \textcolor{black}{($T\times F$)}& 3$\times$3, 3$\times$3, 4$\times$3, 4$\times$3 \\
			Pooling size \textcolor{black}{($T\times F$)}, only encoder& 1$\times$25 (max pooling) \\\wcline{1-2}
		\end{tabular}
	}
	\vspace{-10pt}
\end{table}

\vspace{5pt}
\section{Proposed method}
\vspace{-8pt}
The proposed method exploits the semantic contexts of scenes (e.g., ``home,'' ``office,'' and ``cooking'') to handle the pre-defined scene-agnostic contexts for SED.
As previously mentioned, the conventional SED methods cannot employ undefined (unseen) contexts of scenes because one-hot representations or identifiers of pre-defined scenes are utilized as the prior contexts for effective SED.
To address this problem, the proposed SED method utilizes the semantic embedding of scenes instead of the one-hot representations or their identifiers.
In Fig. \ref{fig:example_propesed_network}, an example of the proposed network is depicted.
The proposed network is divided into three modules: (a) SED module, (b) semantic embedding module, and (c) alignment module.
Module (a) works as a sound event detector, module (b) transforms contexts of scenes into semantic embedding, and module (c) aligns the semantic embedding with an acoustic embedding (representation learning).

The SED module (a) is a general architecture used in conventional SED.
In Fig. \ref{fig:example_propesed_network}, the acoustic features in the time-frequency domain are input to a feature extractor such as CNN or RNN layers.
The output of the feature extractor is input to a time-distributed classifier such as feedforward neural network (FFN) layers.

The semantic embedding module (b) handles the contexts of not only pre-defined (seen) but also unseen scenes.
In this work, to extract the semantic contexts of scenes, we utilize label texts of acoustic scenes, e.g., ``city center'' or ``home,'' using a procedure similar to that in \cite{hu_interspeech2020,Xie_TASLP2021}. 
The label texts of scenes are input to pre-trained language models, then a vector of the semantic embedding $\bm e$~$\in \mathbb{R} ^{E}$ is produced, where the number of dimensions $E$ depends on the pre-trained language models.
By leveraging the pre-trained language models, the proposed SED model can utilize unseen semantic contexts of scenes, i.e., data not used for training, in inference stages.
Note that not only language models but also models for other modalities can be utilized to employ semantic contexts of the other modalities.

The semantic embedding $\bm e$ might not be appropriate for SED since module (b) comprises  the language-model-trained texts.
Hence, in the alignment module (c), the agreement between the semantic and acoustic embedding spaces is maximized.
More specifically, a transformed semantic embedding $\bm l$~$\in \mathbb{R}^{L}$, which is passed through two FFN layers ($\mathbb{R}^{E}\rightarrow\mathbb{R}^{E'}\rightarrow\mathbb{R}^{L}$), and an acoustic embedding $\bm z$~$\in \mathbb{R}^{A}$, which is a bottleneck feature encoded by an autoencoder (AE), are followed by each FFN layer.
Here, the input of the AE is the output of the last CNN layer of the feature extractor in module (a). 
L1 loss between $\bm l'$~$\in \mathbb{R}^{S}$ and $\bm z'$~$\in \mathbb{R}^{S}$ is then calculated, which is passed through each of the FFN layers. 
Finally, the aligned semantic embedding $\bm l$ is concatenated with each sequence of the first layer of the FFN layers before the time-distributed classifier in module (a) to take advantage of the prior semantic contexts of scenes as follows.

In module (c), to optimize the AE similarly to that in\cite{Deshmukh_interspeech2021}, the following mean squared error between the input $x_{t,f}$ and the reconstructed output $\hat x_{t,f}$ is used:

\vspace{-18pt}   
\begin{align}
\hspace*{0pt} {\mathcal L}_{\rm AE} = \frac{1}{TF}\sum_{t=1}^{T}\sum_{f=1}^{F}(x_{t,f}-\hat x_{t,f})^2,
\label{eq:AE_loss}
\vspace{-30pt}
\end{align}
\vspace{-12pt}

\noindent where $F$ is the number of dimensions of the acoustic feature.
To align the semantic embedding $\bm l'$ and acoustic embedding $\bm z'$, L1 loss is used as follows: 

\vspace{-18pt}   
\begin{align}
\hspace*{0pt} {\mathcal L}_{\rm align} = \frac{1}{S}\sum_{s=1}^{S}| {\it l}'_{s}-{\it z}'_{s}|,
\label{eq:AE_loss}
\vspace{-30pt}
\end{align}
\vspace{-12pt}

\noindent where $l'_{s}$ and $z'_{s}$ represent the elements $s$ of the semantic embedding $\bm l'$ and acoustic embedding $\bm z'$, respectively, and $S$ is the number of dimensions of $\bm l'$ and $\bm z'$.
Overall, to optimize the proposed network, the following objective function is used:

\vspace{-16pt}   
\begin{align}
\hspace*{0pt} {\mathcal L}
 = {\mathcal L}_{\rm SED} + \alpha{\mathcal L}_{\rm AE} + \beta{\mathcal L}_{\rm align},
\label{eq:proposed_loss}
\vspace{-20pt}
\end{align}
\vspace{-15pt}

\noindent where hyperparameters $\alpha$ and $\beta$ are tuned on the development set and set to 0.01 and 1.0, respectively.

\vspace{-5pt}
\section{Experiments}
\label{sec:exp}
\vspace{-8pt}
\subsection{Experimental conditions}
\label{sec:condition}
\vspace{-5pt}
We performed experiments to evaluate the performance of the proposed method using the TUT Sound Events 2016 \cite{Mesaros2016TUTDF}, TUT Sound Events 2017 \cite{Mesaros2017}, TUT Acoustic Scenes 2016 \cite{Mesaros2016TUTDF}, and TUT Acoustic Scenes 2017 \cite{Mesaros2017} datasets.
From these datasets, we chose audio clips comprising four scenes, ``home,'' ``residential area'' (TUT Sound Events 2016), ``city center'' (TUT Sound Events 2017, TUT Acoustic Scenes 2017), and ``office'' (TUT Acoustic Scenes 2016), which include 266 min (development set, 192 min; evaluation set, 74 min) of audio.
There were no sound event labels for the scenes ``office'' in TUT Acoustic Scenes 2016 and ``city center'' in TUT Acoustic Scenes 2017. 
We thus manually annotated the audio clips with sound event labels by the procedure described in \cite{Mesaros2016TUTDF,Mesaros2017}.
The resulting audio clips contained 25 types of sound event label \cite{tonami_IEICE2021}.
The event label annotations for our experiments are available in \cite{Imoto_dataset2019_01}.

\textcolor{black}{
As acoustic features, we used 64-dimensional log-mel energies computed for every 40 ms temporal frame with 50\% overlap, where the clip length was 10 s.
The audio clip shorter than 10 s was zero-padded.
The sampling rate was 44.1 kHz.}
This setup is from the baseline system of DCASE2018 Challenge task4 \cite{setting01}.
We performed the experiments using ten initial values of the model parameters. 
A segment-based metric \cite{Mesaros2016_MDPI} was used to evaluate the performance of SED.
We set the segment size to the frame length. 
The threshold value for binarizing $\sigma (y_{n,t})$ was 0.5.
As an optimizer, we used AdaBelief \cite{AdaBelief}.
\textcolor{black}{For AdaBelief, smoothing parameters $\beta_{1}$ and $\beta_{2}$, and a small number $\epsilon$ were tuned on the development set and set to 0.9, 0.999, and $10^{-3}$, respectively.} 
The activation function was Swish  \cite{swish}. 

As baseline models, the conventional methods denoted as $\textless$~$\tt 1$~$\textgreater$ to $\textless$~$\tt 6$~$\textgreater$ in Table \ref{tbl:overall_results} were used.
As one of the baseline models, we used the convolutional neural network and bidirectional gated recurrent unit (CNN--BiGRU) \cite{SED_CRNN}.
In addition, to verify the usefulness of the proposed method, we used a model combining SED and sound activity detection (SAD) based on multitask learning (MTL), referred to as ``MTL of SED \& SAD'' \cite{SED_SAD}, and a model combining SED and ASC, referred to as ``MTL of SED \& ASC'' \cite{tonami_IEICE2021}.
SAD is the process of estimating all active events in a time frame.
The reason for choosing MTL of SED \& SAD is that this modern method, in which no scene information is considered, is simple yet effective. 
MTL of SED \& ASC is multitask-learning-based SED with ASC, which exploits scene labels for ASC.
Moreover, we used Conformer \cite{conformer_miyazaki} as the SED module, which achieved the best performance in DCASE2020 Challenge task4.
To verify the effectiveness of the semantic embedding of scenes, the one-hot representation of scenes was also evaluated as a substitute for their semantic embedding \cite{Komatsu_icassp2020}.

In this work, as the language model in semantic embedding module (b) (Fig. \ref{fig:example_propesed_network}), we used bidirectional encoder representations from transformers (BERT) \cite{BERT_proc} and generative pre-trained transformer 2 (GPT2) \cite{radford2019language}, which are pre-trained models and frozen when the SED module (a) and the alignment module (c) are trained.
For GPT2, we used the last sequence of the layer before the final layer to extract the semantic embedding.
In module (b), $E$ was 768 for BERT or 1280 for GPT2.
In module (c), $E'$, $L$, $A$, and $S$ were 256, 64, 64, and 32, respectively, which were tuned on the development set.
For the AE, we used a CNN--AE.
The scale of the encoder of the CNN--AE was smaller than that of the decoder of the CNN--AE because the output of the CNN layers in module (a) was followed by the CNN--AE. 
Other experimental conditions are listed in Table~\ref{tbl:parameter}, including the parameters of the CNN--AE and the baseline of module (a).
In Table~\ref{tbl:parameter}, X $\times$ Y denotes a filter size of X along the frequency axis by Y along the time axis.

\setcounter{table}{2}
\vspace{-15pt}
\begin{table}[t]
	\small
	\caption{Seen or unseen scenes for each experiment}
	\vspace{0pt}
	\label{tbl:seen_unseen}
	\centering
	%\hspace{40pt}
	\scalebox{0.90}[0.90]{
		\begin{tabular}{lcc}
			\wcline{1-3}
			&&\\[-7pt]
			& Seen & Unseen\\\hline
			&&\\[-9pt]
			\multirow{2}{*}{\textbf{Experiment 1}}& ``city center'' ``home'' & \multirow{2}{*}{-} \\
			&  ``office'' ``residential area'' &\\
			&&\\[-10pt]\hline
			&&\\[-9pt]
			\multirow{4}{*}{\textbf{Experiment 2}}&  ``home'' & ``city center'' \\
			&  ``office'' ``residential area'' & ``downtown''\\
			&&\\[-10pt]\cline{2-3}
			&&\\[-9pt]
			&  ``city center'' ``home'' & ``residential area'' \\
			&  ``office'' & ``apartment''\\\wcline{1-3}
		\end{tabular}
	}
	\vspace{-10pt}
\end{table}

%---------------------------------
\vspace{5pt}
\subsection{Experimental results}
\label{sec:results}
\vspace{-5pt}
%---------------------------------
We conducted the following two experiments.

\noindent
\begin{itemize}
	\item \textbf{[Experiment 1]}: We investigated the overall SED performance of the proposed method and the extent to which the semantic representation of scene contexts is beneficial for SED.
	In the inference stages, as the semantic contexts of the scenes, we employed the scene labels assigned to the training data, i.e., the seen semantic contexts shown in Table \ref{tbl:seen_unseen}.
	
	\item \noindent\textbf{[Experiment 2]}: The aim of the experiments was to demonstrate that unseen semantic contexts boost SED under the audio of unseen scenes.
	We verified the SED performance using SED models learned without the audio and semantic contexts of evaluation target scenes.
	In the inference stages, SED was performed for the audio of the unseen scenes using the unseen contexts  shown in Table \ref{tbl:seen_unseen}.
	
\end{itemize}

\noindent\textbf{[Experiment 1]}
Table \ref{tbl:overall_results} shows the overall results for SED in terms of F-score, where micro and macro denote overall and classwise scores, respectively.
The numbers to the right of $\pm$ indicate standard deviations.
Each method is tagged with an ID from $\textless$~$\tt 1$~$\textgreater$ to $\textless$~$\tt 14$~$\textgreater$, where $\textless$~$\tt 1$~$\textgreater$ to $\textless$~$\tt 6$~$\textgreater$ are the conventional methods and the others are the proposed methods.
``One-hot'' means the one-hot encoding of scenes is input to module (b) as a substitute for BERT or GPT2 embedding.
``Direct concatenation'' indicates that the scene representations (one-hot or BERT or GPT embedding) directly concatenate with the SED module.
``Aligned concatenation'' indicates that the scene representations concatenate with the SED module after the processing of the alignment by module (c). 
The results show that the F-score of the proposed methods is higher than that of the conventional methods.
In particular, $\textless$~$\tt 13$~$\textgreater$ and $\textless$~$\tt 12$~$\textgreater$ improved the micro and macro F-scores by 4.34 and 3.13 percentage points compared with those of $\textless$~$\tt 4$~$\textgreater$ and $\textless$~$\tt 1$~$\textgreater$, respectively.
However, $\textless$~$\tt 7$~$\textgreater$ and $\textless$~$\tt 8$~$\textgreater$ did not improve the F-score compared with that of $\textless$~$\tt 5$~$\textgreater$.
This means that the space of the non-aligned semantic embedding of scenes is different from that of the acoustic embedding of scenes.
To confirm this, the embedding spaces based on $\textless$$\tt 12$$\textgreater$ are visualized by principal component analysis (PCA) in Fig. \ref{fig:visualization}.
As shown in Fig. \ref{fig:visualization} (A), the distance between ``city center'' and ``residential area'' \textcolor{black}{(6.60)} is almost the same as that between ``home'' and ``office'' \textcolor{black}{(6.02)}.
However, in Fig. \ref{fig:visualization} (B), the distance between the clusters of ``city center'' and ``residential area'' is smaller than that between the clusters of ``home'' and ``office.''
In Fig. \ref{fig:visualization} (C), the distance between ``city center'' and ``residential area'' \textcolor{black}{(69.91)} is smaller than that between ``home'' and ``office'' \textcolor{black}{(23.19)}, that is, the embedding $\bm l$ is well aligned with the acoustic feature.
``downtown'' and ``apartment'' will be described later.

\setcounter{figure}{2}
\begin{figure}[t!]
	\centering
	\includegraphics[width=0.99\columnwidth]{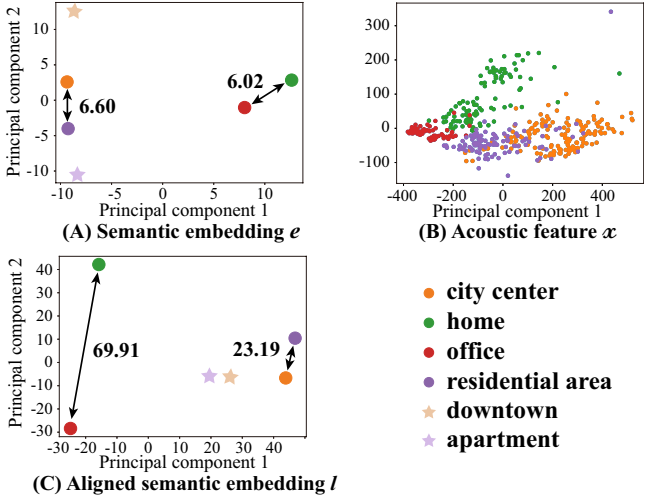}
	\vspace{-10pt}
	\caption{Visualization of space of semantic embedding and acoustic feature}
	\label{fig:visualization}
	\vspace{-15pt}
\end{figure}

\noindent\textbf{[Experiment 2]}
Tables \ref{tbl:unseen_results} (A) and (B) give the SED performance for each event in the audio of the scenes ``city center'' and ``residential area'' using the SED models trained without the audio and contexts of ``city center'' and ``residential area,'' respectively.
In this experiment, the conventional method $\textless$~$\tt 1$~$\textgreater$ and proposed method $\textless$~$\tt 12$~$\textgreater$, which achieved the best macro F-scores, were used.
Here, ``Unseen context'' was input to module (b) in the inference stages for the evaluated unseen scenes.
\textcolor{black}{Note that the conventional methods $\textless$~$\tt 5$~$\textgreater$ and $\textless$~$\tt 6$~$\textgreater$ cannot exploit unseen contexts of scenes.}
To further verify the SED performance using synonyms of the contexts, we added two contexts, ``$\tt downtown$'' and ``$\tt apartment$,'' which are similar to ``$\tt city~center$'' and ``$\tt residential~area$,'' respectively, from \cite{thesaurus}.
Note that there was no corresponding audio for ``$\tt downtown$'' and ``$\tt apartment$,'' as shown in Fig, \ref{fig:visualization}.
Table \ref{tbl:unseen_results} (A) shows the results obtained when the SED models were trained using the audio of scenes ``home,'' ``office,'' and ``residential area'' with the corresponding semantic contexts, i.e., scene labels ``$\tt home$,'' ``$\tt office$,'' and ``$\tt residential~area$.''
The audio and contexts of ``$\tt city~center$'' and ``$\tt downtown$'' were not used for the training.
The results indicate that the unseen semantic contexts of scenes are effective for more accurate SED under the audio of unseen scenes.
Using the unseen context ``$\tt city~center$,'' the events are well detected compared with method $\textless$~$\tt 1$~$\textgreater$.
In other words, the unseen context indeed boost SED under the audio of the unseen scene.
This is because the audio and context of ``city center'' are similar to those of ``residential area''  used for training, as shown in Fig. \ref{fig:visualization}.
Using the unseen context ``$\tt downtown$,'' which is similar to ``$\tt city~center$,'' as shown in Fig. \ref{fig:visualization} (C), most of the events are well detected compared with method $\textless$~$\tt 1$~$\textgreater$.
Thus, the coarse semantic contexts (synonyms) of scenes are also useful for SED even when the fine context is not known.
Table \ref{tbl:unseen_results} (B) shows the results obtained when the SED models were trained using the audio of scenes ``city center,'' ``home,'' and ``office'' with the corresponding semantic contexts, i.e., scene labels ``$\tt city~center$,'' ``$\tt home$,'' and ``$\tt office$.''
Using the unseen context ``$\tt apartment$,'' the SED performance is highly degraded compared with method $\textless$~$\tt 1$~$\textgreater$.
This implies that the SED performance of some events is sensitive to even the slightest non-audio-aligned semantic embedding.

\setcounter{table}{3}
\begin{table}[]
	\centering
	\caption{SED results for each event and scene, obtained using SED models learned without audio and semantic contexts of evaluation target scenes. $\textless$$\tt 1$$\textgreater$ is a conventional  method and $\textless$$\tt 12$$\textgreater$ is a proposed method. }
	\vspace{5pt}
	\label{tbl:unseen_results}
	\scalebox{0.86}[0.86]{
		\small
		\begin{tabular}{p{10.4mm}m{31mm}ccc}
			\multicolumn{5}{c}{\large (A) evaluated scene: city center} \\[0pt]
			\multicolumn{5}{c}{trained scenes: home, office, residential area} \\[0pt]
			\wcline{1-5}
			& & & &\\[-8pt]
			\multirow{2}{*}{Method} & \multirow{1}{*}{Unseen} & \multirow{2}{*}{car} & \multirow{2}{*}{children} & large \\
			& \multirow{1}{*}{context} &  &  & vehicle\\\hline
			& & & &\\[-9pt]
			$\textless$$\tt 1$$\textgreater$ & - & 41.72\% & 3.17\%  & 14.59\% \\\hdashline
			& & & &\\[-9pt]
			$\textless$$\tt 12$$\textgreater$ & \tt city center& \bf 42.93\% & \bf 4.68\%  & \bf 15.88\%\\[-2pt]
			& & & &\\[-9pt]
			\multirow{1}{*}{$\textless$$\tt 12$$\textgreater$} & \tt downtown & 40.29\% &  3.97\% & 15.35\% \\\wcline{1-5}
			& & & &\\[2pt]
			\multicolumn{5}{c}{\large (B) evaluated scene: residential area} \\[0pt]
			\multicolumn{5}{c}{trained scenes: city center, home, office} \\[0pt]
			\wcline{1-5}
			& & & &\\[-8pt]
			\multirow{2}{*}{Method} & \multirow{1}{*}{Unseen} & \multirow{2}{*}{car} & brakes & people\\
			&  \multirow{1}{*}{context} &  & squeaking & talking\\\hline
			& & & &\\[-9pt]
			$\textless$$\tt 1$$\textgreater$ & - & \bf 41.41\% & 0.00\% &3.09\% \\\hdashline
			& & & &\\[-9pt]
			\multirow{1}{*}{$\textless$$\tt 12$$\textgreater$} & \tt residential area & \multirow{1}{*}{36.80\%} &  \multirow{1}{*}{\bf 0.68\%}  & \bf \multirow{1}{*}{5.56\%} \\[-2pt]
			& & & & \\[-9pt]
			\multirow{1}{*}{$\textless$$\tt 12$$\textgreater$} & \tt apartment & 24.09\% &  0.00\% & \textcolor{black}{4.52\%} \\\wcline{1-5}
			& & & &\\[-9pt]
		\end{tabular}
	}
	\vspace{-10pt}
\end{table}

\vspace{-8pt}
\section{Conclusion}
\label{sec:conc}
\vspace{-8pt}
In this paper, we proposed scene-informed SED where pre-defined scene-agnostic contexts are available.
In the proposed method, pre-trained large-scale language models were used, which enables SED models to employ contexts of not only pre-defined (seen) scenes but also unseen scenes.
We further investigated the extent to which the semantic representation of scene contexts is helpful for SED.
The experimental results show that the proposed method improves the micro and macro F-scores by 4.34 and 3.13 percentage points compared with the conventional Conformer- and CNN--BiGRU-based SED, respectively.
Moreover, we confirmed that the unseen semantic contexts of the scenes can boost SED under the audio of unseen scenes.
In our future work, we will investigate the SED performance guided by the semantic contexts of not only scenes but also sound events. 

%---------------------------------

%---------------------------------

\vspace{-5pt}
\section{Acknowledgement}
\label{sec:ack}
\vspace{-8pt}
This work was supported by JSPS KAKENHI Grant Number JP19K20304.
\vspace{-8pt}

%---------------------------------

%\vspace{20pt}
\clearpage
\bibliographystyle{IEEEbib}
\bibliography{IEEEabrv,refs_et_al}
\end{document}